\documentclass[11pt,preprintnumbers,nofootinbib,onecolumn]{revtex4-1}
\pdfoutput=1

\usepackage{color}
\usepackage{graphicx}
\usepackage{amsmath}
\usepackage{amssymb}

\usepackage[colorlinks]{hyperref}
\hypersetup{allcolors=[rgb]{0,0,1}}
\newcommand{\orcid}[1]{\href{https://orcid.org/#1}{#1}}
\usepackage{scalerel}
\usepackage{graphicx}
\usepackage[normalem]{ulem}
\usepackage{subcaption}
\usepackage{lineno}

\def\lsim{\raise0.3ex\hbox{$\;<$\kern-0.75em\raise-1.1ex
\hbox{$\sim\;$}}}
\def\gsim{\raise0.3ex\hbox{$\;>$\kern-0.75em\raise-1.1ex
\hbox{$\sim\;$}}}

\keywords{water-based liquid scintillator (WbLS), LAPPD, neutrino telescope, Yemilab}

\begin{document}

\vspace*{-3cm}
\title{ Neutrino Telescope at Yemilab, Korea }

\author{Seon-Hee~Seo}
\email{sunny.seo@ibs.re.kr}
\thanks{\orcid{0000-0002-1496-624X}}
\affiliation{Center for Underground Physics, Institute for Basic Science, 55 Expo-ro Yuseong-gu, Daejeon 34126, Korea \\ }

\vglue 1.6cm

\vspace*{2.cm}

\begin{abstract}
A new underground lab, Yemilab, is being constructed in Handuk iron mine, Korea. 
The default design of Yemilab includes a space for a future neutrino experiment.
We propose to build a water-based liquid scintillator (WbLS) detector of 4$\sim$5 kiloton size at the Yemilab. 
The WbLS technology combines the benefits from both water and liquid scintillator (LS) in a single detector
so that low energy physics and rare event searches can have higher sensitivities due to the larger size detector with increased light yield. 
No experiment has ever used a WbLS technology since it still needs some R\&D studies, as currently being performed by THEIA group. 
If this technology works successfully with kiloton scale detector at Yemilab then it can be applied to future T2HKK (Hyper-K 2$^{nd}$ detector in Korea)
to improve its physics potentials especially in  the low energy region. 

\end{abstract}

\date{March 13, 2019}


\maketitle


\section{Introduction}
\label{intro}

Over several decades water Cherenkov detectors have been very successful in neutrino physics as proven by Kamiokande (3 kiloton water)~\cite{KAMIOKANDE}, Super-Kamiokande (50 kiloton water)~\cite{SK} and SNO (1 kiloton D$_{2}$O)~\cite{SNO} experiments resulting in Nobel prizes for the first observation of neutrinos from Supernova burst (SN1987A) and for the discovery of neutrino flavor transformations. Based on these success, the next generation Hyper-Kamiokande experiment (2 x 260 kiloton water Cherenkov detector) is being pursued and soon one of the two detectors is very likely to be built in Japan~\cite{HK}. The possibility of building the other detector in Korea~\cite{HKK}  is currently being discussed, benefiting from a much longer baseline for the J-PARC neutrino beam. \\

Liquid scintillator (LS) detectors have been also very successful in neutrino physics from the first detection of neutrinos in 1956 by Reines and Cowan team leading to Nobel prize in 1995, to more recent KamLAND~\cite{KL}, CHOOZ~\cite{CHOOZ}, Double CHOOZ~\cite{DC}, RENO~\cite{RENO} and Daya Bay~\cite{DB}. The largest LS neutrino detector, JUNO~\cite{JUNO}, is also being built in China to determine neutrino mass ordering as well as to measure neutrino oscillation parameters very precisely and to perform neutrino astrophysics. \\

At IBS, 
we are interested in developing this new technology, combining benefits of both water Cherenkov and LS detectors. the water-based liquid scintillator (WbLS) detector, so that the performance is better than either of the two types of detectors. Water has longer absorption length and also gives directional information via Cherenkov light, while LS gives high light yield but no directional information. 
The concept of WbLS and its application to neutrino physics is well described in reference ~\cite{theia}.  The THEIA group lead by USA teams has been working on developing this kind of an advanced detector, WbLS, but it still at the level of R\&D. If this new technology is successfully developed then a WbLS detector could achieve better physics results than water Cherenkov or LS detector alone especially at low energy physics and rare event searches. \\

\begin{figure}
\begin{center}
\includegraphics[width=7cm]{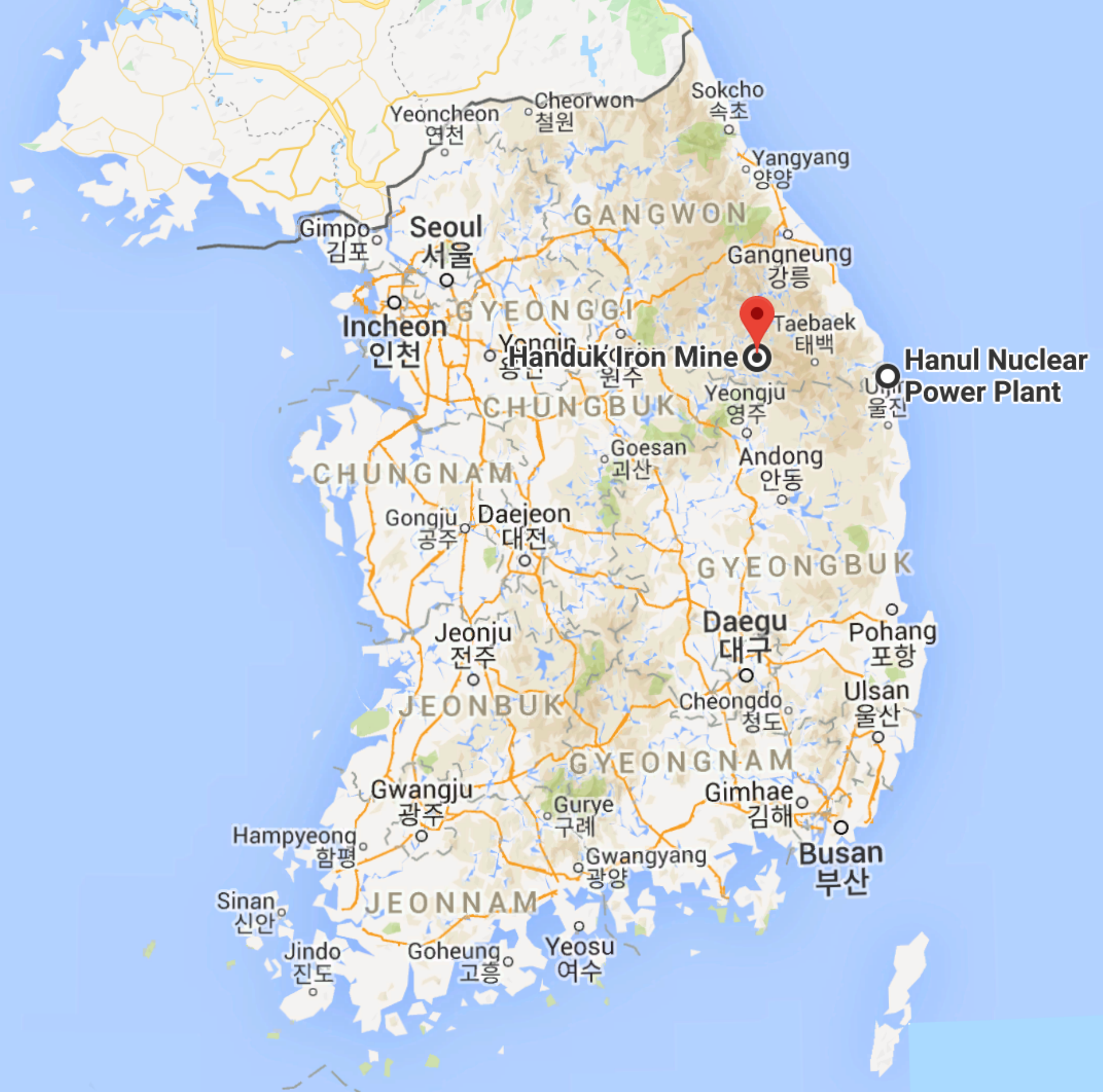}
\includegraphics[width=9cm]{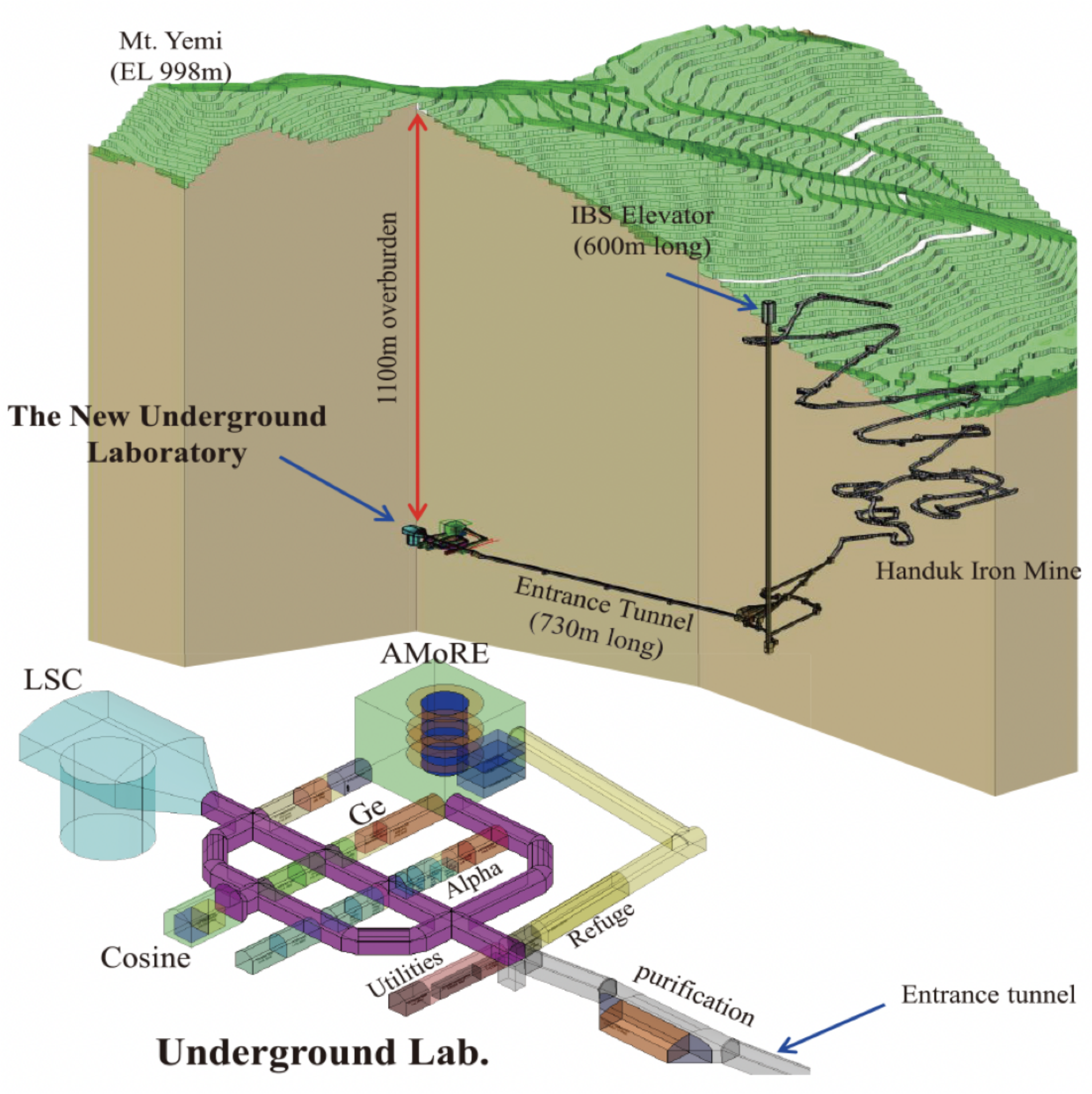}
\end{center}
\caption{
(Top) Handuk iron mine location (lattitude: 37.188639 deg, longitude: 128.659406 deg) in Jeongseon, Korea.
There are three commercial reactor complex along the east cost and the closest one (Hanul Nuclear Power Plant) is at about 65~km. 
(Bottom) The new underground Yemilab ($\sim$1000~m overburden) to be completed in early 2021 at the Handuk iron mine. 
A space for a neutrino experiment is allocated in the baseline design. 
Tunnel length is about 730~m from the entrance tunnel reached by a shaft (600~m vertical distance). 
}
\label{f:yemilab}
\end{figure}

By early 2021 a new underground laboratory called Yemilab ($\sim$1000~m overburden) will be ready at the Handuk iron mine in Jeongseon, Gangwon province, Korea (see Fig.~\ref{f:yemilab}). A space for a neutrino experiment is allocated in the current design of the Yemilab
and we propose to use this space for a WbLS detector (4$\sim$5 kiloton) after successful R\&D phase. 
To make this possible we would like to work closely together and/or collaborate with the THEIA group. 
The 4$\sim$5 kiloton WbLS detector at Yemilab, is expected to produce interesting (astro-)physics results. 
It will be indeed the first neutrino telescope in Korea based on a new technology. 
The best application of this technology would be for T2HKK to be discussed later. 

\section{WbLS: New technology}

\subsection{WbLS R\&D}

As described already both water and liquid scintillator (LS) are good medium for neutrino detection and well proven technologies. The detectors based on each of these medium have been very successful but no experiment exists using the mixture of the two medium which would overcome short-comings of each medium by complimenting one another. \\

\begin{figure}
\begin{center}
\includegraphics[width=12cm]{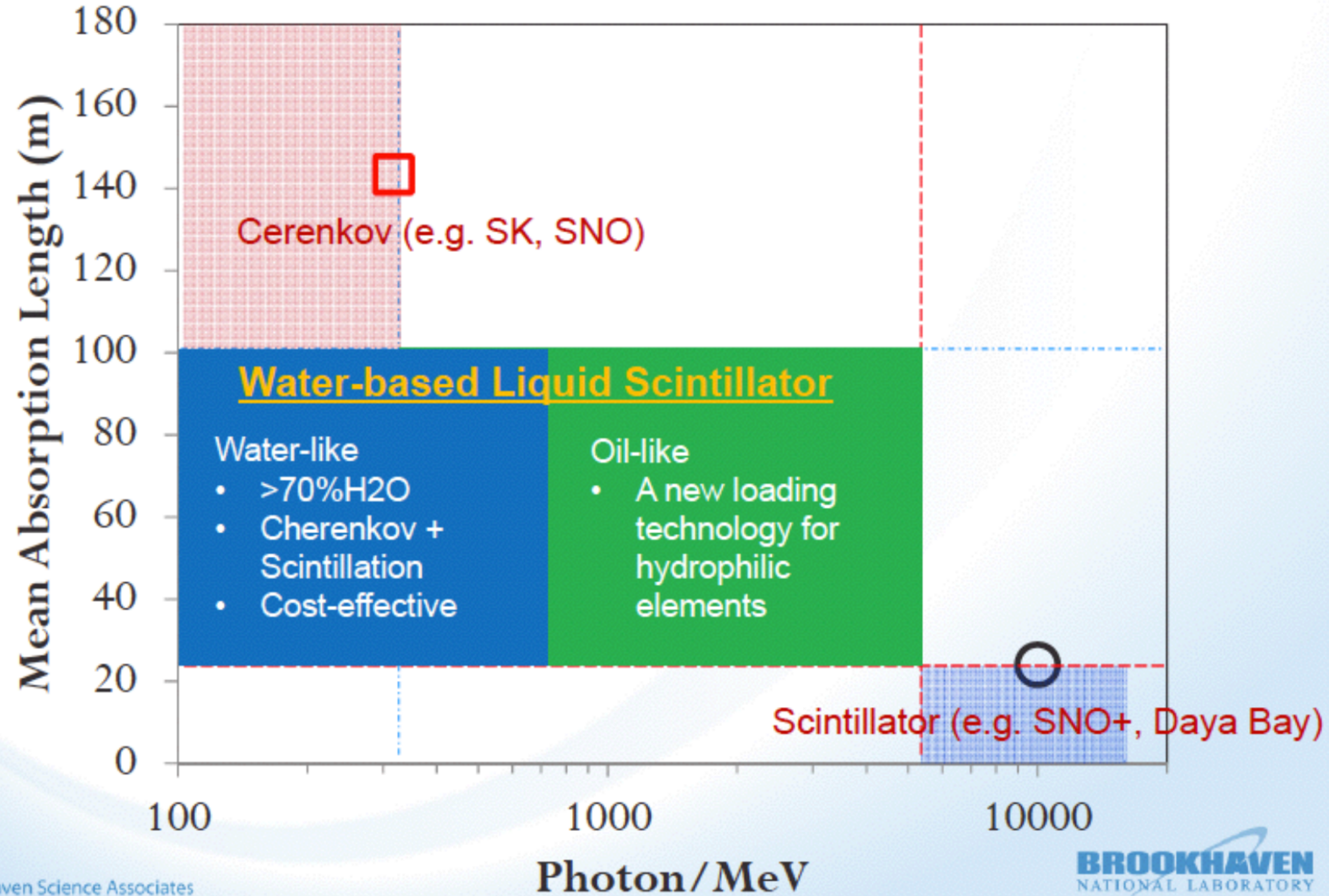}
\end{center}
\caption{
Mean absorption length (m) vs light yield (photon/MeV) for water and LS. Taken from M. Yeh's talk slide in 2014. 
}
\label{f:comp}
\end{figure}

Figure~\ref{f:comp} shows characteristics of the two medium in mean absorption length (m) vs light yield (photons/MeV).  Water has much larger ($>$ 100~m) absorption length than LS ( $< \sim$20~m), which allows us to build a larger detector. In addition to this, water gives a Cherenkov signal giving directional information which cannot be obtained in LS detector due to isotropic nature of scintillation light. On the other hand, LS has much larger ($\sim$10,000 photons/MeV) light yield than water ($\sim$200 photons/MeV) and this allows us to explore energies below Cherenkov threshold and to also have better energy resolution. Because of these benefits of the two medium it is very good to combine the two in one detector to compensate and compliment each other. 
Depending on the fraction of LS that is added to water, WbLS is classified as water-like below 30\% whereas above its oil-like. 
In this research only water-like WbLS is discussed; mainly due to its cost-effectiveness, environment-friendliness, and easy of loading additional isotopes. \\

The largest LS detector so far was KamLAND (1 kiloton LS) in Japan but the near future JUNO in China will be the biggest one (20 kiloton LS) once it is completed around 2021. In principle this could be the biggest LS detector due to the limitation of the short attenuation length of LS. However, with WbLS technology, it is possible to build much bigger detector so that more interesting and broader physics program can be pursued, from particle physics (proton decay, neutrinoless double beta decay, etc) to multi messenger astro-particle physics using neutrinos from the Sun, Supernova, or GW etc. \\

To improve the discovery potential of proton decay one could add Gadolinium (Gd) for neutron tagging to suppress main background from atmospheric neutrinos. Therefore we need R\&D for the Gd loading option for proton decay search as well as reactor neutrino physics and supernova relic neutrino search. The other alternatives are Li and B for neutron tagging.  For a neutrinoless double beta decay search Te, Nd, or Xe isotopes needed to be loaded in WbLS. Loading these isotopes are easier thanks to their hydrophilic nature but long term stability and light yields also need to be studied. \\

The following studies need to be performed for WbLS R\&D~\cite{theia}. 

\begin{itemize}
\item{Light yield}
\item{Light loss due to absorption and scattering (attenuation)}
\item{Fraction of Cherenkov to scintillation light produced}
\item{Time profile of the emitted scintillation light (for Cherenkov/scintillation separation, and particle ID)}
\item{Effect of isotope loading on each of these for different fractions of isotope loading}
\item{Stability of each of these properties over time}
\end{itemize}

\subsection{Fast timing resolution photo sensor R\&D}

\begin{figure}
\begin{center}
\includegraphics[width=12cm]{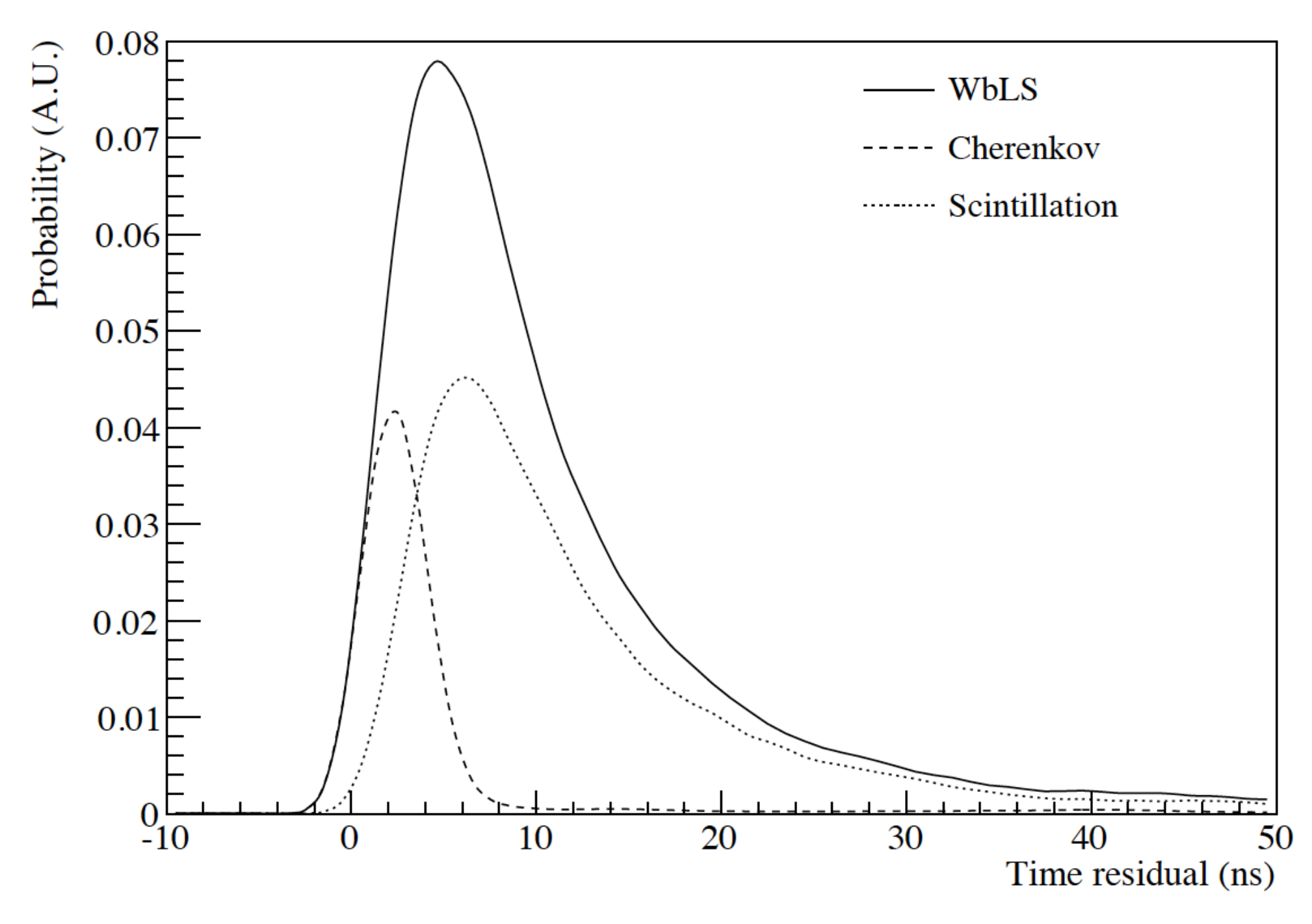}
\end{center}
\caption{
Time profile of PMT hits from Cherenkov and scintillation light, and their sum, modeled as a 1\%
mixture of pseudocumene in water, which reproduces the factor of four increase in intrinsic light yield THEIA group has measured for the WbLS cocktail. 
(``A.U.'' are arbitrary units for the probability scale). Taken from reference ~\cite{theia}.
}
\label{f:time}
\end{figure}

To obtain the full benefits from the WbLS technology, the capability to separate Cherenkov and 
scintillation light signals is necessary. Directional information from Cherenkov light signal will help to suppress backgrounds.
Figure~\ref{f:time} shows the time profile of observed photons in a WbLS detector with an optical model of 1\% pseudocumene in water, where the smearing due to the PMT transit time spread is included. About 60\% of photons in the first 4 ns are Cherenkov photons and this fraction is reduced with increased LS fraction. \\

Hamamatsu H11780 (12 inch) would be a reasonable off-the-shelf choice for photo sensors of WbLS detector since they have reasonable timing resolution ($\sim$1.2~ns FWHM) to separate Cherenkov and scintillator light and also they have high quantum efficiency ($\sim$34\%). To reduce cost of photo-coverage, reflectors or wavelength-shifting plates could be also used with small effects on timing, however this needs to be studied. \\

\begin{figure}
\begin{center}
\includegraphics[width=12cm]{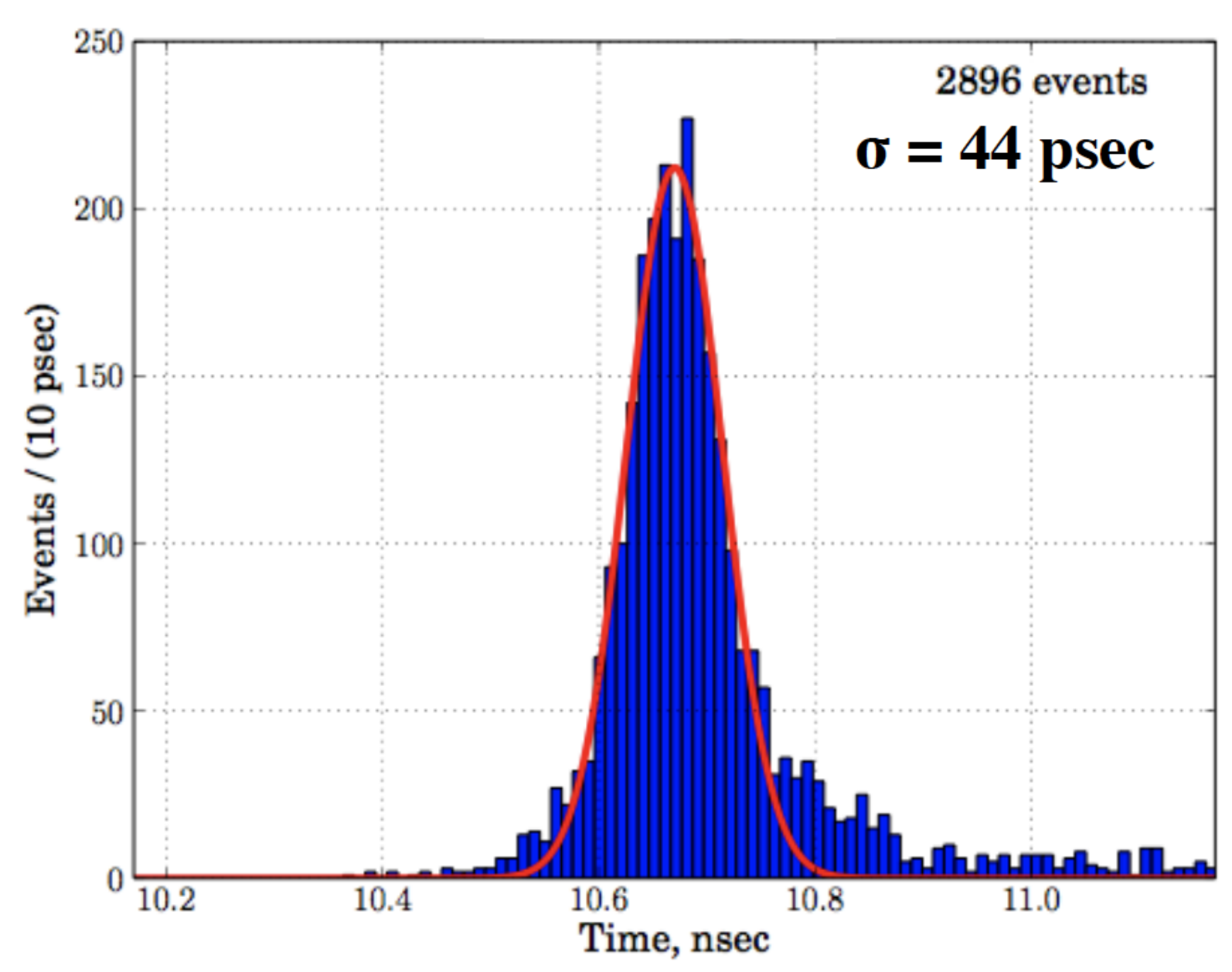}
\end{center}
\caption{
Preliminary result on the transit-time spread (TTS) of a pair of 8'' x  8'' Microchannel plates, measured from
an LAPPD designed 8'' anode. Taken from reference ~\cite{theia}. 
}
\label{f:lappd}
\end{figure}

LAPPDs (Large Area Pico-second Photo Detectors) using MCP (micro-channel plate) technology could provide with an excellent timing resolution ($\sim$60~ps) as shown in Fig.~\ref{f:lappd}. Unlike PMTs LAPPDs are imaging tubes which can identify the position and time of the single incident photons in an individual sensor. Its high spatial resolution ($<$ 1~cm) makes it possible to increase fiducial volume of the WbLS detector due to improved vertex resolution near the detector boundary where most events exist. However, LAPPDs are still in R\&D stage and being tested. 
ANNIE (Accelerator Neutrino Neutron Interaction Experiment)~\cite{ANNIE} in Fermilab would be the first running experiment (26 ton water Cherenkov detector with Gd loading) 
to test some LAPPDs using PSEC4 readout system in their phase II run, very soon. 
Our group would like to contribute to developing LAPPD technology to be used in WbLS detector even though our baseline WbLS detector design is to use off-the-shelf high QE PMTs which are well understood and characterized. 

\section{Science Programs}
\label{l:program}

The new WbLS technology being developed is expected to revolutionize neutrino experiments by combining water and LS in one detector. The WbLS technology is environment friendly, cost-effective for large detection volume required for future neutrino experiments, and at the same time accessing low energy below Cherenkov threshold. This novel technique allows us to have directional information which is a powerful tool to reduce background from signals coming from sources such as Sun, GRBs, Supernova bursts, GW etc. or to reduce charged-current (CC) background. 
Very good timing resolution photo sensors such as LAPPDs will allow us to separate out Cherenkov light from scintillation light, 
so that directional information is extracted to suppress background. \\

After successful R\&D on WbLS and possible R\&D on fast timing resolution photo sensors with a fast readout system, a 4$\sim$5 kiloton WbLS detector would be built at Yemilab. 
The following science programs will be pursued using data taken with the 4$\sim$5 kiloton WbLS detector, and sensitivity studies need to be done. 

\begin{figure}
\begin{center}
\includegraphics[width=13.5cm]{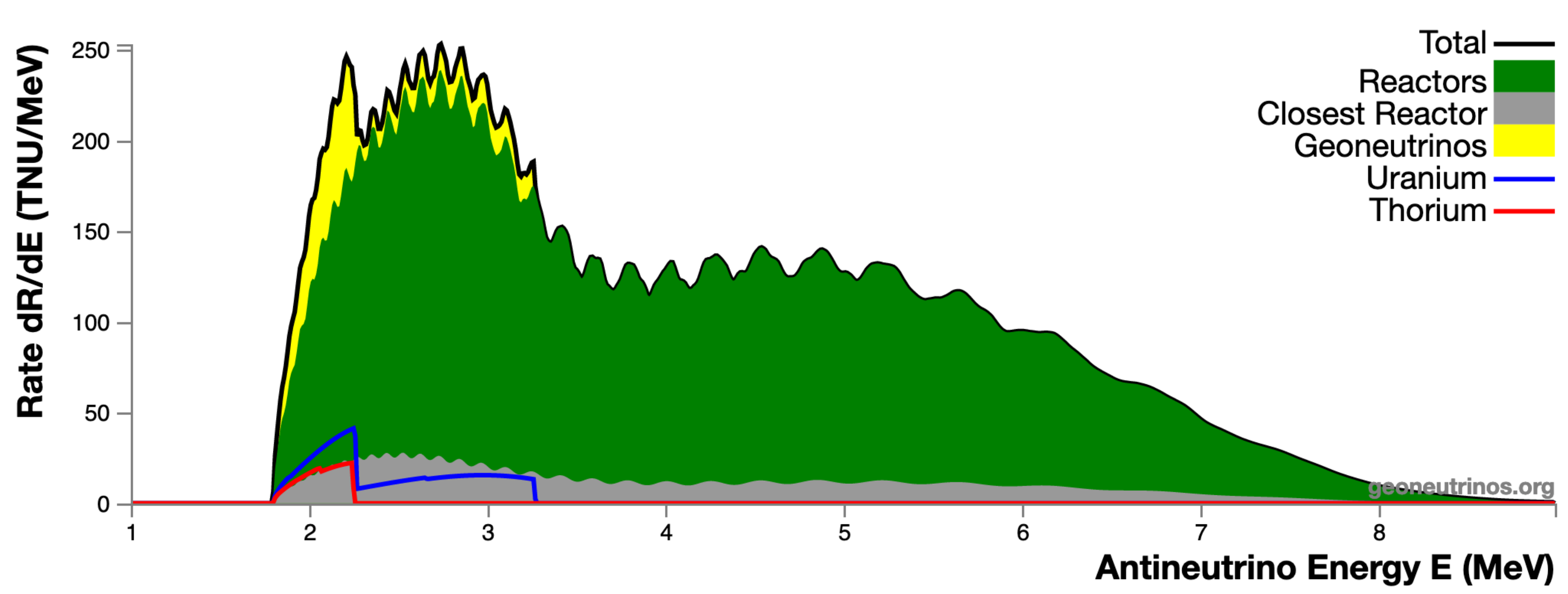}
\end{center}
\caption{
Reactor and geo neutrino event rates in Yemilab~\cite{antiNu_web}. 
}
\label{f:reactorNu}
\end{figure}

\begin{itemize}
\item{Solar neutrino physics: \\
 Upturn study; Day/Night asymmetry study to resolve $\delta m^{2}_{21}$ conflict between Super-K/SNO vs KamLAND; CNO neutrino detection etc.
}
\item{Proton decay search: \\
 Very good sensitivities are expected especially in p$\rightarrow \nu K^{+}$ and $n \rightarrow 3 \nu$ channels. 
}
\item{Detection of Supernova burst neutrinos: \\
 About 80$\sim$120 events are expected to be observed form a SN burst in 10~kpc~\footnote{\label{theia_scale} This number is scaled by detector size from reference ~\cite{theia}.}.
 Multi-messenger astronomy (Supernova Early Warning System)
}
\item{Supernova Relic Neutrino (SRN) search: \\
 About 5 events are expected for 10 years~\footnote{This number is scaled by detector size from reference ~\cite{theia}.}. 
}
\item{Indirect dark matter search: \\
  Need to set sensitivities. 
}
\item{Detection of neutrinos from GW sources: \\
 This is challenging but upper limits can be set. 
}
\item{Reactor neutrino physics: \\
About 2\,000 reactor neutrinos per year for 5 kiloton WbLS is expected assuming 100\% detection efficiency and neutrino oscillation 
from all reactors in Korea (see Fig.~\ref{f:reactorNu}).
The closest Hanul reactor complex is at about 65~km away (see Fig.~\ref{f:yemilab}) and it dominantly contributes to the reactor neutrino flux.
}
\item{Geo neutrino physics: \\
 Explore interior (tomography) of the Earth. About 20 geo neutrinos per year for 5 kiloton WbLS is expected assuming 100\% detection efficiency (see Fig.~\ref{f:reactorNu}).
}
\end{itemize}

\begin{figure}
\begin{center}
\includegraphics[width=12cm]{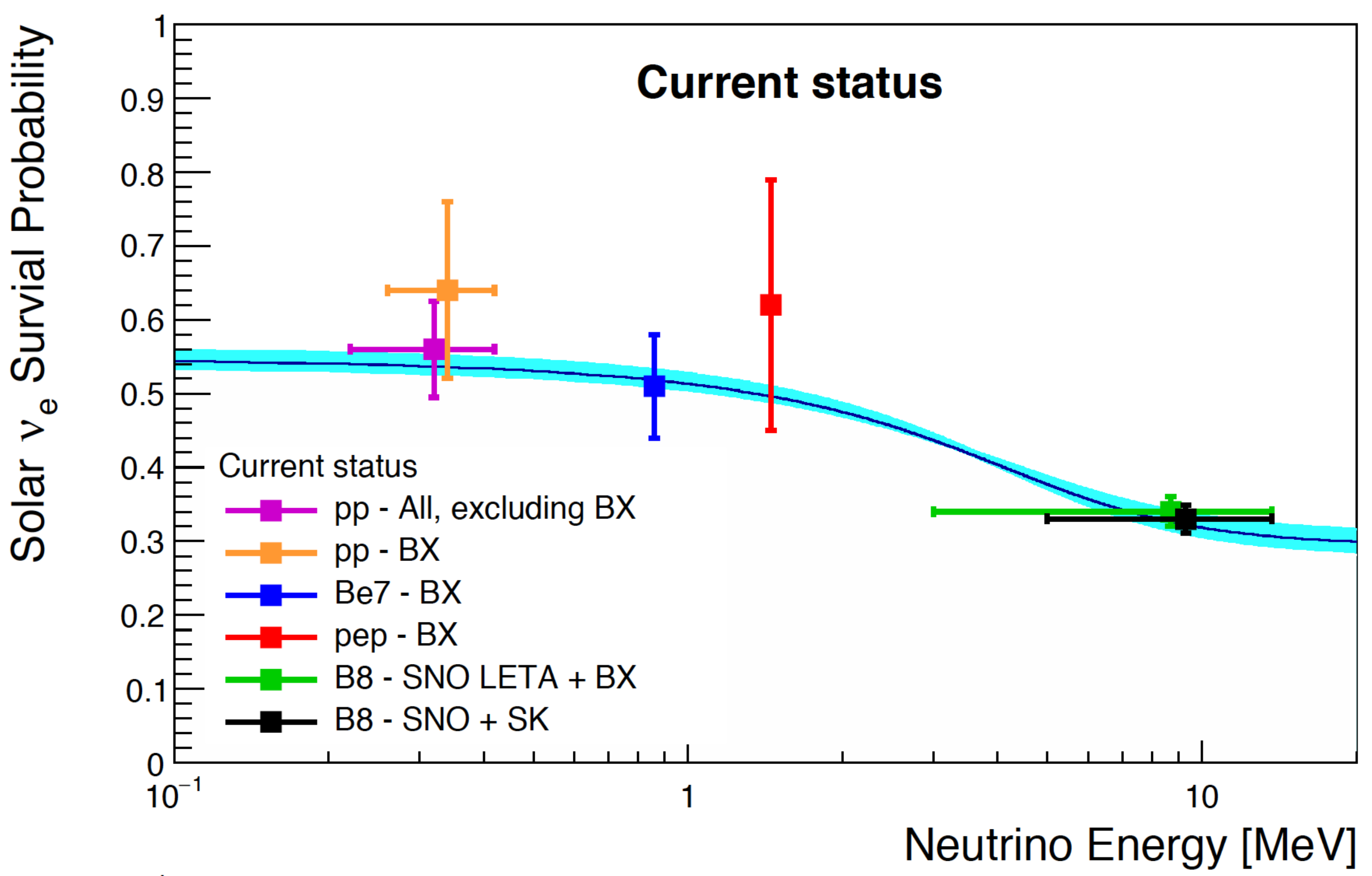}
\end{center}
\caption{ Solar neutrino survival probability vs. neutrino energy in MeV. Squares with error bars represent solar neutrino fluxes from current measurements by Borexino, SNO and SK. A cyan band represents expected solar neutrino survival probability from standard solar neutrino model with MSW effect. With 4$\sim$5 kiloton WbLS detector at Yemilab it might be possible to reduce the uncertainties to the level of the expected one (cyan band). 
}
\label{f:solarNu}
\end{figure}

Among the science programs above, solar neutrino physics sensitivity would be achieved very well with a few kiloton scale purified detector at Yemilab. We expect to reduce currently measured solar neutrino flux uncertainties (see the squares with error bars in Fig.~\ref{f:solarNu}) to the level of model uncertainty (see the cyan band in Fig.~\ref{f:solarNu}). With the small uncertainties of the solar neutrino flux measurements one can nail down whether standard solar MSW-LMA model is correct or there is new physics such as mass varying neutrinos, non-standard neutrino interaction, sterile neutrinos, etc. 
Current tension (2 $\sigma$ level) of $\delta m^{2}_{21}$ between KamLAND and SK/SNO can be studied by the kiloton scale WbLS detector through Day/Night asymmetry measurement. 
Understanding this tension is very important in the current paradyme of 3 neutrino oscillation to rule out any new physics or to discover new physics,
and therefore adding more data to this observation is important~\cite{solar_dm2}. 
Discovery of CNO neutrinos might be also possible and this would resolve solar metallicity problem important for nuclear astrophysics and stellar evolution. \\

There has been no neutrino telescope so far in Korea but the 4$\sim$5 kiloton WbLS detector would be a good 1$^{st}$ neutrino telescope in Korea in the near future 
to carry out multi messenger neutrino physics.

\section{The Yemilab}

Yemilab is a new underground facility owned by IBS
and it takes about 2.5 hours by car to get there from IBS HQ in Daejeon where CUP (Center for Underground Physics) is.
The closest reactor facility (Hanul Nuclear Power Plant) is at Uljin about 65~km away and can produce $\sim$15~GW$_{th}$ power.
Handuk iron mine in Gangwon province is the only iron mine active in Korea and kindly agreed to provide an underground space for Yemilab. 
A shaft ride of 600~m vertical distance is needed to reach to the entrance tunnel of the Yemilab. 
Tunnel length is about 730~m with 12 degree downward slope from the entrance. \\

Yemilab is born mainly for CUP to place their future upgraded AMoRE $0\nu\beta\beta$ and COSINE dark matter search experiments due to lack of space at current Yang Yang underground lab (700~m overburden) in Sockcho city
where AMoRE prototype detector has recently finished data taking and COSINE-100 is still taking data.
There is also a space at Yemilab for a future neutrino experiment of up to 5 kiloton target size. 
A closed school nearby the Handuk mine will be converted to a surface lab for Yemilab. \\

To carry out our research, first of all, three laboratory rooms are needed in IBS HQ building or the surface lab to perform R\&D for WbLS (15~m x 15~m), photo sensors (15~m x 15~m), and electronics (10~m x 10~m). For WbLS R\&D, liquid purification, clean room, and fume duct facilities are needed as well as acrylic vessels, liquid scintillator, chemistry lab equipment etc.  For photo sensor R\&D, dark room (with a temperature control capability) and clean room are needed as well as equipment like oscilloscopes, FADC, TDC, HV, computers etc. For electronics R\&D, a clean room is needed as well as equipment for front-end electronics and DAQ test etc. \\

At Yemilab the pit (D: 20~m x H: 20~m) for a future neutrino experiment would be a nice home for our WbLS detector. Other than this, we need electronics and monitoring rooms where air conditioners are equipped to keep constant temperature and humidity for front-end/DAQ electronics and computers. We also need a space to assemble our detector components in Yemilab. 
The key research equipment and facilities needed in our research are summarized in the Table~\ref{t:facil}. 
Note that the sizes of the lab rooms and details of facilities and equipment might be changed later. 

\begin{table}
\begin{center}
\scalebox{0.75}{
\begin{tabular}{|c|c|c|c|c|}\hline
                  &   WbLS R\&D     &  Photo sensor R\&D    &  Electronics R\&D  & Yemilab  \\ 
                  & at IBS HQ       &  at IBS HQ            &  at IBS HQ         &          \\   \hline\hline
                  &   &   &  & Electronics room  \\
Lab rooms         & 15~m x 15~m & 15~m x 15~m & 10~m x 10~m & (4~m x 4~m), \\ 
                  &               &              &             & Monitroing room \\ 
                  & & & & (4~m x 4~m) \\ \hline
                  & Liquid purification  & Dark room  &  &  Area to assemble \\ 
Facilities        & facility, clean room,   & w/ temp. control,  & Clean room  &  detector \\
                  & fume ducts, etc   &  clean room &  &   components \\ \hline
                  & Acrylic vessels,  &  Oscilloscopes, & Front-end electronics, &   \\ 
Equipment         &  LS,  &  FACDs, TDCs, HVs, & DAQ R\&D and test &   \\
                  &  chemistry lab  & PMTs, LAPPDs, &  equipment, etc. &   \\
                  & equipments, etc. & computers, etc. & & \\ \hline
\end{tabular}
}
\caption{Necessary facilities and equipment at IBS HQ building or the surface lab, and Yemilab.}
\label{t:facil}
\end{center}
\end{table}

\section{Future Application of WbLS to T2HKK}

WbLS technology has a huge potential for future neutrino physics since it allows to build a very large detector providing with an excellent capability in low energy physics 
as well as high energy physics. 
With an enlarged pit of 50~m (D) x 50~m (H) at Yemilab, 50 kiloton WbLS detector can be built and it is expected to produce very competitive physics results 
for the science programs listed in section ~\ref{l:program}. 
The WbLS technology can be also applied to the future Korean Neutrino Observatory (KNO, Hyper-K 2$^{nd}$ detector in Korea, a.k.a. T2HKK, 260 kiloton water Cherenkov detector) 
where many Korean (astro-)physicists/astronomers are currently putting efforts together due to its great physics potentials which will be further improved by WbLS. 
Very interesting physics studies can be additionally done as listed below. 
\begin{itemize}
\item{Measurement of CP violation in leptonic sector}
\item{Neutrino mass ordering determination}
\item{Flavor textures of mass eigenstates}
\item{Study of non-standard neutrino interaction}
\end{itemize}
Impact on beam neutrino physics with WbLS, however, is still needed to be studied such as e-$\mu$ separation power and Cherenkov ring counting efficiency, etc. 
The WbLS detector in this research (4$\sim$5 kiloton at Yemilab) would be considered as a good demonstrator for KNO/T2HKK. \\

Note that USA’s future flagship neutrino experiment, DUNE, uses liquid Argon (LAr) technology, Japan’s future flagship neutrino experiment, Hyper-K 1$^{st}$ detector, 
uses water Cherenkov technology, and China’s future flagship neutrino experiment, JUNO, uses LS technology. Korea’s future flagship neutrino experiment, if KNO/T2HKK ever happens, 
could use WbLS technology so that our research could provide an excellent platform for such a future. \\

WATCHMAN (WATer CHerenkov Monitor for ANtineutrinos)~\cite{WATCHMAN} collaboration between USA and UK groups also considers to use in the future a WbLS based detector 
with fast photo sensors for nuclear security and fundamental neutrino phsycis programs by studying neutrinos from reactors, the Earth, the Sun, and the distant stars.

\section{Conclusion}

WbLS is a new technology to be used in a future neutrino experiment at an intensity frontier. 
Synergy between water and LS in a single large detector would make low energy physics and rare event seaerches more plausable than water detector alone
by pushing down energy threshold and improving energy resolution by the added LS. 
We propose to perform R\&D study on WbLS in Korea and then to build a neutrino telescope (4$\sim$5 kiloton WbLS) at a new Yemilab ($\sim$1000~m overburden) 
currently being constructed in Korea, where upgraded AMoRE and COSINE experiments will be also placed. 
To maximize the potential of WbLS technology, fast timing resolution photo sensors like LAPPDs are needed.
Physics programs are as broad as water Cherenkov detectoror or LS detector, from astro-physics to particle physics. \\

A bigger detector gives better physics sensitivities and therefore if this few kiloton scale WbLS is successful 
then this technology can be applied to a even bigger detector such as T2HKK (260 kiloton), the best application of the WbLS technology. 
The few kiloton WbLS detector, indeed, could be the first neutrino telescope in Korea for multi-messenger (astro-)physics as well as particle physics, 
and we believe it is very likely to happen thanks to the new underground Yemilab to be completed by early 2021.

\section{Acknowledgements}

This work was supported by the National Research Foundation of Korea (NRF) grant funded by the Korea government (MSIT)
(No. 2017R1A2B4012757 and IBS-R016-D1-2018-b01).



\begin{thebibliography}{99}

\bibitem{KAMIOKANDE} K. Hirata et al. (Kamiokande-II Collaboration), Phys. Rev. Lett. {\bf 58}, 1490 (1987).
\bibitem{SK} Y. Fukuda et al. (Super-Kamiokande Collaboration), Phys. Rev. Lett. {\bf 81}, 1562 (1998).
\bibitem{SNO} Q. R. Ahmad et al. (SNO Collaboration), Phys. Rev. Lett. {\bf 87}, 071301 (2001).
\bibitem{HK} K.~Abe {\it et al.} [Hyper-Kamiokande Collaboration],
  arXiv:1805.04163 [physics.ins-det].
\bibitem{HKK} K.~Abe {\it et al.} [Hyper-Kamiokande Proto- Collaboration],
  PTEP {\bf 2015}, 053C02 (2015)
  doi:10.1093/ptep/ptv061
  [arXiv:1502.05199 [hep-ex]].
\bibitem{KL} K. Eguchi et al. (KamLAND Collaboration), Phys. Rev. Lett. {\bf 90}, 021802 (2003).
\bibitem{CHOOZ} M. Apollonio et al. (CHOOZ Collaboration), Phys. Lett. {\bf B466}, 415 (1999).
\bibitem{DC} Y. Abe et al. (Double Chooz Collaboration), Phys. Rev. Lett. {\bf 108}, 131801 (2012).
\bibitem{RENO} J. K. Ahn et al. (RENO Collaboration), Phys. Rev. Lett. {\bf 108}, 191802 (2012). 
\bibitem{DB} F. P. An et al. (Daya Bay Collaboration), Phys. Rev. Lett. {\bf 108}, 171803 (2012).
\bibitem{JUNO} F. An et al., J. Phys. G 43 (2016) 030401; arXiv:1507.05613 (2015).
\bibitem{theia} J. R. Alonso et al., arXiv:1409.5864.
\bibitem{ANNIE} http://annie.fnal.gov/
\bibitem{antiNu_web} Barna, A.M. and Dye, S.T., ``Web Application for Modeling Global Antineutrinos'' arXiv:1510.05633 (2015). 
\bibitem{RENO_PRL} J. K. Ahn et al., Physical Review Letters {\bf 108} (2012) 191802.

\bibitem{jinping_solar} J. F. Beacom et al., Chinese Physics C Vol. {\bf 41}, No. 2 (2017) 023002; arXiv:1602.01733.
\bibitem{t2hkk} K. Abe et al. (Hyper-Kamiokande proto-Collaboration), PTEP 2018.
\bibitem{solar_dm2} S.~H.~Seo and S.~J.~Parke,
  Phys.\ Rev.\ D {\bf 99}, no. 3, 033012 (2019)
  doi:10.1103/PhysRevD.99.033012
  [arXiv:1808.09150 [hep-ex]].

\bibitem{WATCHMAN} https://neutrinos.llnl.gov/projects/ait-watchman

\end{thebibliography}
\end{document}